# Skin Lesion Segmentation and Classification for ISIC 2018 by Combining Deep CNN and Handcrafted Features

Redha Ali, Russell C. Hardie, Manawaduge Supun De Silva, and Temesguen Messay Kebede

Signal and Image Processing Lab
Department of Electrical and Computer Engineering
University of Dayton, 300 College Park, Dayton OH 45469-0232
Email: almahdir1@udayton.edu

## Abstract

This short report describes our submission to the ISIC 2018 Challenge in Skin Lesion Analysis Towards Melanoma Detection [1] for Task1 and Task 3. This work has been accomplished by a team of researchers at the University of Dayton Signal and Image Processing Lab. Our proposed approach is computationally efficient are combines information from both deep learning and handcrafted features. For Task3, we form a new type of image features, called hybrid features, which has stronger discrimination ability than single method features. These features are utilized as inputs to a decision-making model that is based on a multiclass Support Vector Machine (SVM) classifier. The proposed technique is evaluated on online validation databases. Our score was 0.841 with SVM classifier on the validation dataset.

## 1  Introduction

The ISIC 2018 Skin Lesion Analysis Towards Melanoma Detection [1] is broken into three separate tasks: 1) Lesion Segmentation, 2), Lesion Attribute Detection and 3) Disease Classification. This report addresses Task 1 and Task 3 which are Lesion Segmentation and Disease Classification.  The goal of the Task 3 is to classify skin lesion images into 7 Possible disease categories – Melanoma (MEL), Melanocytic nevus (NV), Basal cell carcinoma (BCC), Actinic keratosis intraepithelial carcinoma (AKIEC), Benign keratosis (BKL), Dermatofibroma (DF), and Vascular lesion (VASC) [1]. Participants are being ranked using a normalized multi-class accuracy metric (balanced across categories) [1].

## 2  Task 1 Methodology

We proposed a hybrid Deep and Handcrafted system for lesion segmentation. Our hybrid technique makes use of the following two methods: Deep CNN and Gaussian mixture models (GMMs) [3]. Both of These are described below.

### 2.1 Deep and efficient UNet

We implemented an elegant architecture, the so-called "fully convolutional network" [2]. The network architecture illustrated in Fig. 1. The pattern in our segmentation networks requires the downsampling of an image between convolutional and ReLU layers, and then upsample the output to match the input size. The network starts with an image "Input Layer," which describes the image size that the network can process. All training and testing images resized to [224 224

3]. The Down sampling network starts with the 3x3 convolution, Batch normalization layer and followed by a rectified linear unit ReLU layers. At the end of each downsample stage, there is 2x2 max pooling operation with stride 2. After stage 2 in downsampling, we double the number of feature channels. The upsampling is done using the transposed convolution layer commonly referred to as "deconv" or "deconvolution" layer. Every step in the upsampling of the feature map followed by a 3x3 convolution, Batch normalization and followed by a ReLU. The cropping is significant due to the loss of border pixels in every convolution. At the end of the network, we have a 1x1 convolution used to map the feature vector to the number of classes. In total the network has 109 layers.

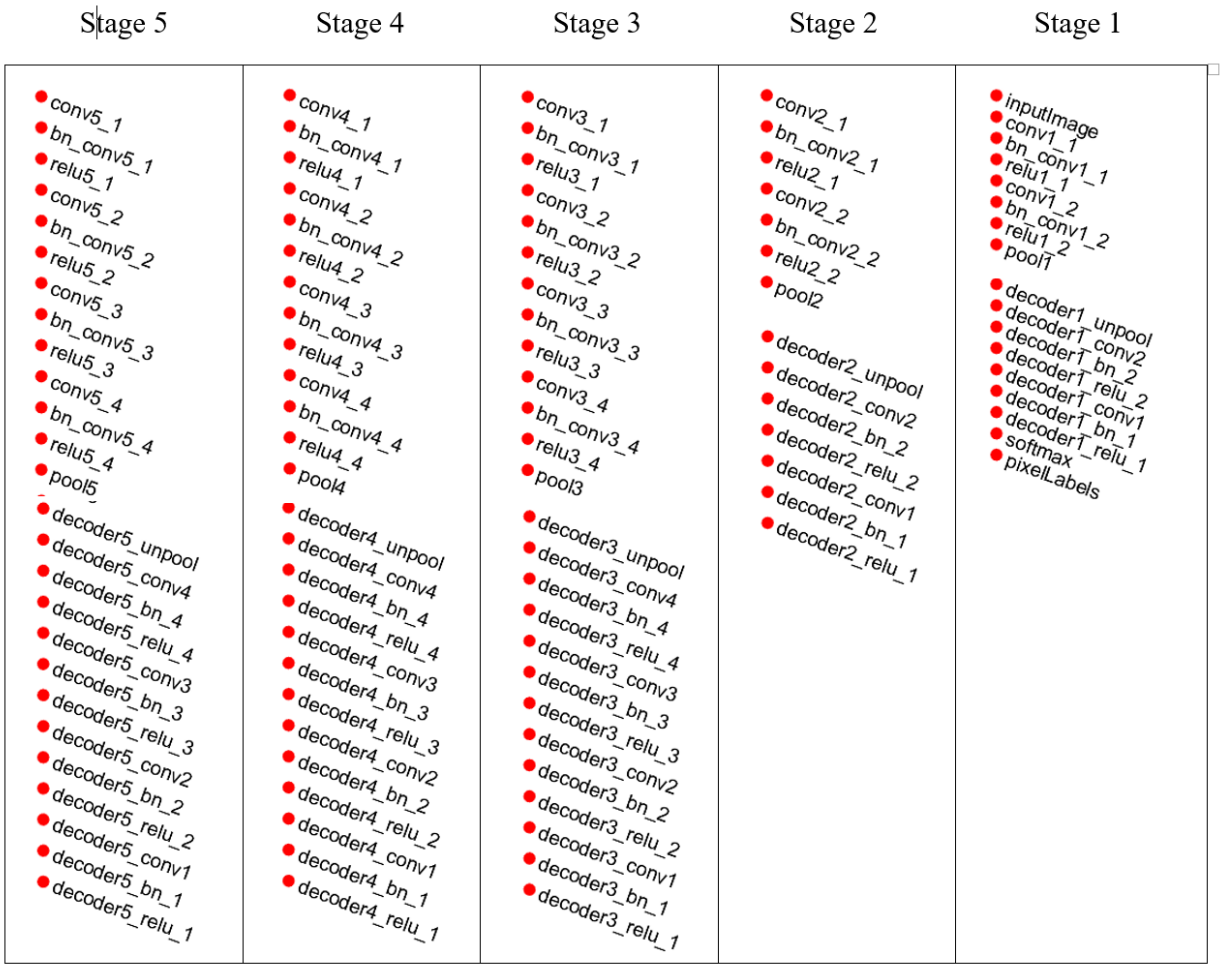

Fig. 1: U-net architecture where each red circle corresponds to a layer. The name of layers is denoted on the side of the circle.

**2.2 Gaussian mixture models (GMMs)**

This model uses the probability density functions of the tissue types to distinguish lesion tissue from normal skin tissue [3]. Our team has submitted this approach in separate submissions using traditional classifiers with hand-crafted features.

To fuse the two system together and obtain the final results. We come up with the best threshold that switches to select between a traditional segmentation approach [3] and deep learning method. This switch is based on estimated lesion area. Since the GMM seems better for smaller lesions and Unet better for larger ones. Therefore, If the lesion area of UNet is smaller than 4508 when the image size is 224 x224, we choose the GMMs mask. Otherwise, select the UNet model result as shown in the Fig. 2.

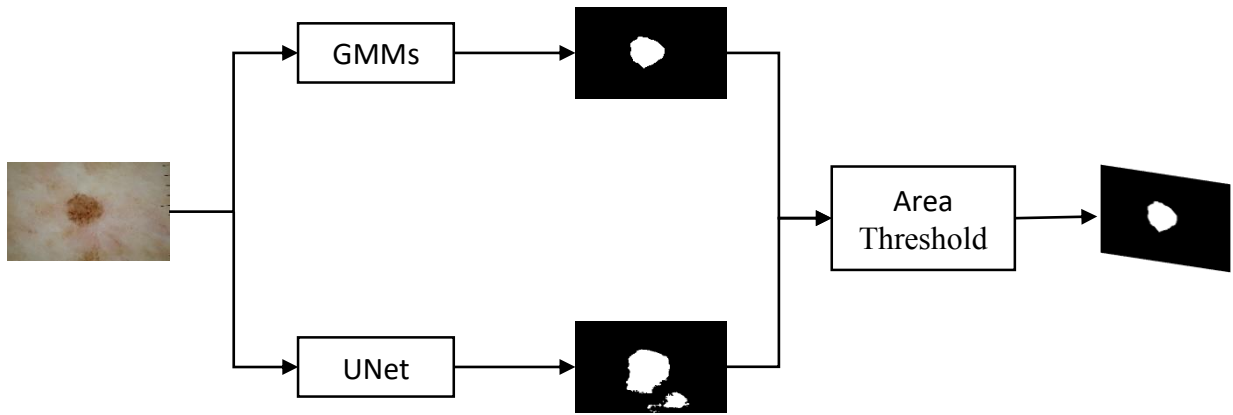

Fig. 2: Shows a hybrid Deep and Handcrafted system

## 3. Task 3 Methodology

Fig. 3 illustrates our proposed classification approach. First, we have trained two convolutional neural networks (CNN) on the available training data for Lesion Diagnosis. The training data consists of 10015 images. All training examples have been resampled to 244 x 224 x 3. We have trained out CNN's without any data augmentations or geometrically transformed such as rotation, translation, scaling and flipping. Second the 200 handcrafted feature [3]. The features are computed from the RGB image with respect to the lesion segmentation. Finally, the handcrafted features [3] are concatenated with CNN features to form the final feature vector. This final feature set is fed to a decision-making model based on multiclass support vector machine (SVM) classifier.

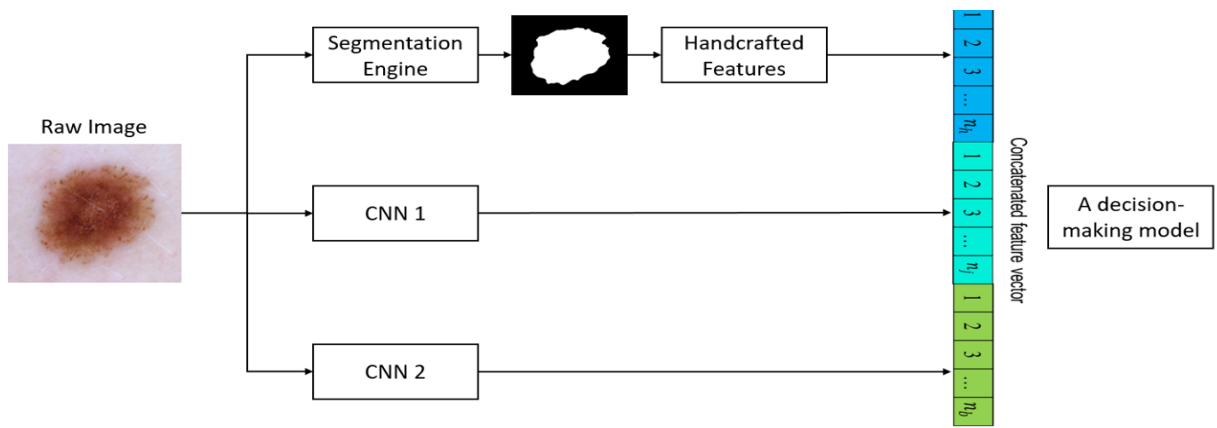

**Fig. 3:** Shows Proposed classification system.

## 4. Experimental Results

### 4.1 Task 1

The results for Task 1 have been obtained using the provided validation dataset. The validation scores are for our information and are not proposed to be made public. We have tested our model on the provided validation data (193 samples in total). The mean overlap score on the testing data is 0.735.

### 4.2 Task 3

In this Task, we employ the ISIC 2018 validation dataset [1] for system evaluation. The results show that our model obtains promising performance with class averaged recall of 0.841.

## 5. Conclusion

This research proposes a robust system for lesion segmentation and disease from dermoscopy images, which offers the vision of achieving an improved and more accurate classification of lesions from images. Our proposed method is based on the use of hybrid features, which are a combination of handcrafted features and deep learning features. Hybrid features provide richer information than that obtained using a single feature extraction method. Hybrid features significantly enhance the segmentation in Task 1 and classification Task 3 accuracy compared to the use of a single method.